\documentstyle[11pt,psfig]{article}
\def \xb {{\bf{x}}}

\begin{document}

\title{Interfacing Interpreted and Compiled \\
Languages to Support Applications on a \\
Massively Parallel Network of Workstations (MP-NOW)
\thanks{This research was supported by Sarnoff Corporation internal R\&D,
DARPA grant F30602-96-CO297 and NSF grant AST~93-15368.}}
\author{Jeremy Kepner$^{a,c}$, Maya Gokhale$^{b,d}$, Ron Minnich$^{b,d}$, \\
Aaron Marks$^b$ and John DeGood$^b$ \\
$^a$Princeton University, Princeton, NJ \\
$^b$Sarnoff Corporation, Princeton, NJ \\
$^c$Current Address: MIT Lincoln Laboratory, Lexington, MA \\
$^d$Current Address: Los Alamos National Laboratory, Los Alamos, NM
}
\maketitle

\begin{abstract}

  The advent of Massively Parallel Network of Workstations (MP-NOW)
represents an important trend in high performance computing.  The rise
of interpreted languages (e.g.  Visual Basic, MATLAB, IDL, Maple and
Mathematica) for algorithm development, prototyping, data analysis and
graphical user interfaces (GUIs) represents an important trend in
software engineering.  However, using interpreted languages on a MP-NOW
is a significant challenge.  We present a specific example of a very
simple, but generic solution to this problem.  Our example uses an
interpreted language to set up a calculation and then interfaces with a
computational kernel written in a compiled language (e.g., C, C++,
Fortran).  The interpreted language calls the computational kernel as an
external library.  We have added to the computational kernel an
additional layer, which manages multiple copies of the kernel running on
a MP-NOW and returns the results back to the interpreted layer.  Our
implementation uses The Next generation Taskbag (TNT) library developed
at Sarnoff to provide an efficient means for implementing task
parallelism.  A test problem (taken from Astronomy) has been implemented
on the Sarnoff Cyclone computer which consists of 160 heterogeneous
nodes connected by a ``fat'' tree 100 Mb/s switched Ethernet running the
RedHat Linux and FreeBSD operating systems.  Our first results in this
ongoing project have demonstrated the feasibility of this approach and
produced speedups of greater than 50 on 60 processors.  \end{abstract}

\section{Introduction}

  The increasing power of cheap computers have made Massively Parallel
Network of Workstations (MP-NOW) with super computing capability a
reality (e.g., Beowulf \cite{Ridge97}, Shrimp \cite{Blumrich98}, and
Avalon \cite{Avalon}).  These machines have a variety of desirable
features in comparison to Massively Parallel Processors (MPP) such as
the Cray T3E, SGI Origin 2000, HP Exemplar and IBM SP.  MP-NOW systems
are readily available, offer the flexibility of both commodity
hardware and software, and deliver superior price/performance in terms
of \$/Megaflop.  All of these features along with competitive pressures
on MPP system vendors suggest that MP-NOWs may be the future of high
performance computing. 

  The MP-NOW concept has been under steady development for almost 15
years, beginning with early work at the National Security Agency
\cite{Mock84,Minnich93} and Fermi Labs.
In recent years, the Beowulf project has provided the impetus for
several MP-NOW systems (see \cite{Beowulf}).  One
implementation, theHIVE (the Highly-parallel Integrated Virtual
Environment), consists of a 50 node cluster in a ``fat'' tree network
configuration rather than the typical homogeneous configuration
provided by Ethernet or switched Ethernet.  The root of the tree is
the connection to the world.  The tree has two levels: 10 ``router''
nodes and 40 ``data'' nodes.

  Another MP-NOW is the SHRIMP (Scalable High-performance Really
Inexpensive Multi-Processor) project, which investigates how to
construct high-performance servers (see \cite{Shrimp}).  So far, four
kinds of Shrimp multicomputers have been developed.  One of them uses 16
Pentium PCs as the computing nodes and the Intel Paragon routing network
as the network and implements a custom network interface.  The other
three use Pentium PCs, Pentium SMPs and Pentium-Pro SMPs as compute
nodes, and use a Myricom routing network and network interfaces. 

 Perhaps one of the largest examples of a MP-NOW machine is the the
Sarnoff Cyclone, which is a jointly funded project between DARPA and the
Sarnoff Corporation.  The Cyclone was built to test operating systems
and systems software concepts related to deployment of hundreds to
thousands of independent processes cooperating to solve a particular
task.  It consists of 160 heterogeneous nodes running Unix connected by
a ``fat'' tree interconnect network (Gigabit root, 100 Megabit switched
internal nodes).  The peak performance for this cluster is approximately
24 Gigaflops at a cost of \$10/Megaflop. 

  Increased power on the desktop has also spawned the wider use of
interpreted languages for algorithm development, prototyping, data
analysis and graphical user interfaces (GUIs).  In fact, within certain
scientific and engineering disciplines the use of interpreted languages
has become the accepted standard for applications development (e.g.,
signal processing with MATLAB and satellite image processing with IDL). 
Several efforts are underway to merge these two trends to allow
interpreted languages take advantage of parallel computers.  These
efforts take a variety of approaches from porting an interpreted
language engine \cite{Mercury} to
providing tools which automatically translate the interpreted language
into a parallelizable form of a compiled language \cite{HeteroRT}. 

  In spite of their widespread use in technically challenging fields,
there are many types of calculations that do not run efficiently in
interpreted languages.  Interpreted languages tend to pass data to
subroutines via copying, which allows some of the more powerful
constructs of these languages.  However, creating many duplicates of
large arrays can be prohibitive.  Interpreted languages also are very
inefficient at handling large loop operations, and programmers are
highly encouraged to vectorize their programs to avoid loops.  To avoid
these limitations most interpreted languages provide simple methods for
linking in compiled subroutines (written in C, C++ or Fortran).  This
allows the programmer to take full advantage of an interpreted language
for developing the parts of an application that are generally most time
consuming in a compiled language (e.g., GUI, I/O and graphics) while
still being able to use a compiled language where performance
constraints require it.  This simple model for enhancing the performance
of interpreted languages is widely used and has the advantage of being
highly portable. 

  Our approach to using interpreted languages on a MP-NOW is to take
advantage of the ability of most interpreted languages to call
externally compiled subroutines.  We break our application into an
interpreted layer and a compiled layer.  In the interpreted layer all
the ``code intensive'' operations are performed (i.e., GUI, I/O, program
management) while in the compiled layer resides a computational kernel
which does the CPU intensive component.  Within the compiled layer we
also insert the tools necessary to manage multiple instances of the
computational kernel on a MP-NOW. 

  Many tools have been developed for scheduling processes on a MP-NOW
(e.g., Load Sharing Facility \cite{LSF} and DQS \cite{DQS}), which
provide various features for starting and monitoring distributed tasks. 
Complete meta-computing environments have also been developed (e.g.,
Globus \cite{Globus} and Condor \cite{Condor}) which encompass
distributing computing along with collaborative tools. Each of these
systems can dispatch, checkpoint, and migrate arbitrary binary object
programs.

 To support applications which can be broken down into multiple
independents tasks, Sarnoff has designed and developed a task management
framework for clusters: The Next generation Taskbag (TNT) based on Poet
\cite{Durant97}.  TNT is a client/server based Applications Programming
Interface (API) for distributing and managing multiple tasks on a
MP-NOW.  TNT shares many of the desirable features found in the
aforementioned systems (e.g., fault tolerance, load balancing, job
monitoring, and job rescheduling). The most notable difference is that
the TNT library is fundamentally a simple framework for task
parallelism, and thus is designed to be compiled directly into the
distributed application and does not require any system wide resources
or installation.  This capability is essential for allowing interpreted
languages to take advantage of task based parallelism through externally
compiled computational kernels.

  In the rest of this paper we present a specific demonstration of our
overall approach.  Section two describes the Sarnoff Cyclone MP-NOW. 
Section three presents the details of the interpreted language we used
for this project (IDL: Interactive Data Language).  Section four talks
about the API and other features of the TNT library.  Section five
describes some of the applications that this approach lends itself to,
with particular emphasis on problems in Astronomy and Astrophysics.
Section six describes the implementation of our test problem.  Section
seven gives the parallel scaling results.  Section eight gives our
conclusions and plans for further work. 

\section{The Sarnoff Cyclone MP-NOW}

\begin{figure}[tbh]
\centerline{\psfig{figure=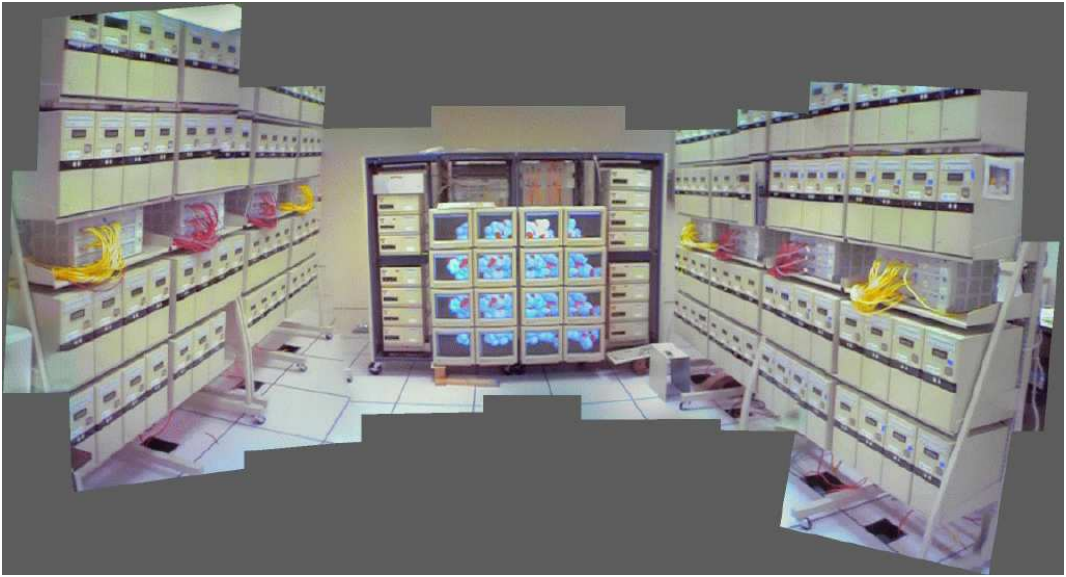,height=3.0in}}
\caption{ {\bf Sarnoff Cyclone MP-NOW} There are a total of 160 heterogeneous
nodes in the cluster: 128 dual 200 MHz Intel P55C (Pentium) workstations, each
with 64 MBytes of RAM and 3 GBytes of disk; 16 533 MHz Digital Alphas (not shown
in image), each with 128 MBytes of RAM and 6 GBytes of disk; 13 90 MHz P54C
workstations, each with 32 MBytes of RAM and 500 MBytes of disk; 2 200 MHz P55C
workstations, each with 32 MBytes of RAM and 500 MBytes of disk; and a dual
450 MHz Pentium II workstation, with 64 MBytes of RAM and 4 GBytes of disk.
There are two networks connecting the nodes of the 128 node dual Pentium CPU
cluster: the public (yellow) and private (red) (see
Figure~\ref{fig:cyclone_arch} for details). Additionally, there is a 4x4 array
of monitors which can present a composite 4000x4000 pixel image.  The monitors
are controlled by the nodes shown on either side.  }
\label{fig:cyclone}
\end{figure}

  The Sarnoff Cyclone is a large, heterogeneous workstation cluster. 
This cluster has been used for a variety of computational tasks, from
real-time MPEG-2 software encoding to distributed World Wide Web
queries.  Figure~\ref{fig:cyclone} provides a mosaic image of the
current Cyclone cluster. 

\begin{figure}[tbh]
\centerline{\psfig{figure=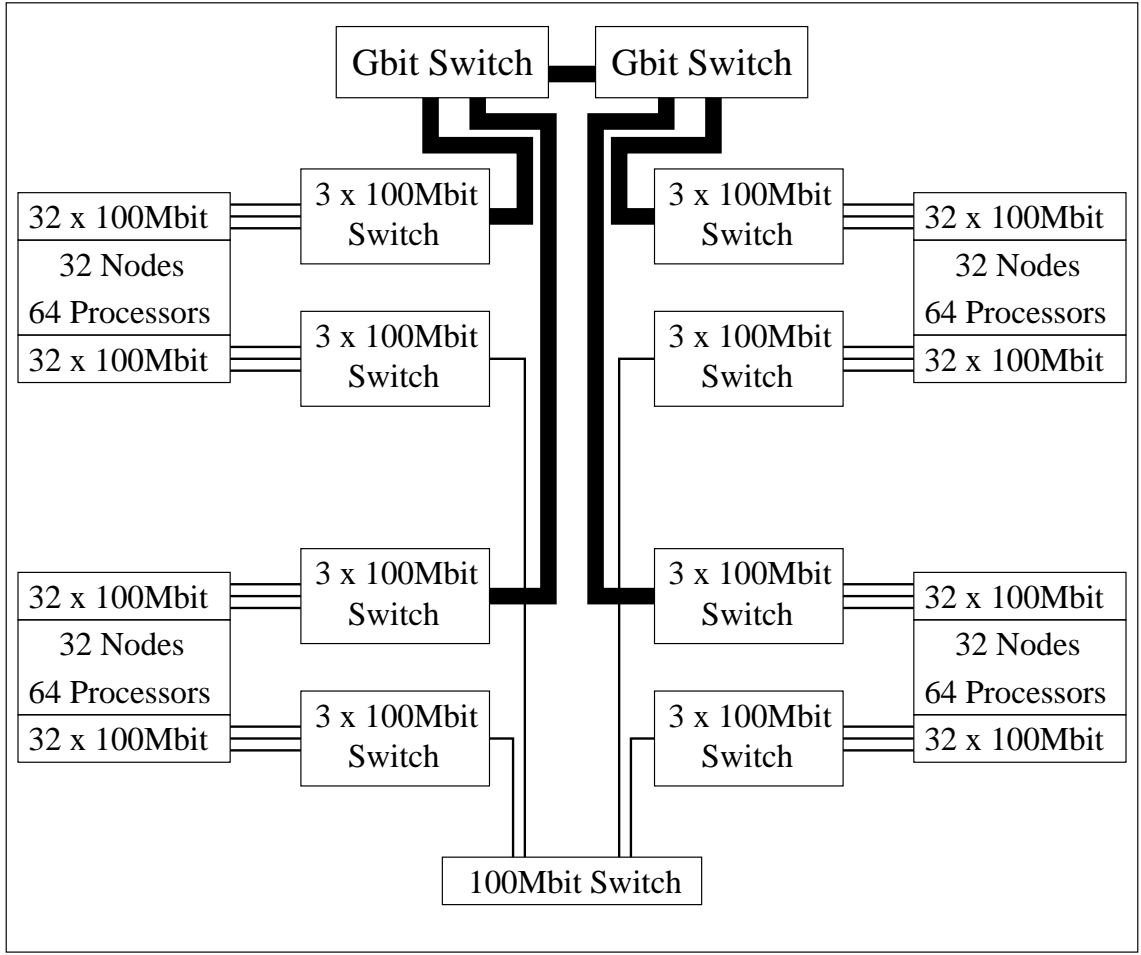,height=5.0in}}
\caption{ {\bf Cyclone Architecture} Topology of the Cyclone cluster.  The
public network (shown in bold) is a ``fat'' tree interconnect with two Gigabit
switches at the root and 12 100 Mbit switches at the nodes.  The 100 Mbit
switches are connected to the Gigabit switches via fiber-optic cable; the PCs
are connected to the 100 Mbit switches via CAT 5 twisted-pair cable.
The private network contains 12 100 Mbit switches all connected to a
single 12-port 100 Mbit switch.
All connections in this network are CAT 5 twisted-pair cable.}
\label{fig:cyclone_arch}

\end{figure}

  Specifically, there are a total of 160 nodes in the cluster: 128 Dual
CPU machines, and 32 Single CPU machines.  Each dual CPU node has 2 200
MHz Intel P55C (Pentium) processors, 64 MBytes of RAM and 3 GBytes of
disk.  Sixteen of the single CPU machines have 533 MHz Digital Alphas,
64 MBytes of RAM and 6 GBytes of disk.  Finally, most of the remaining
16 single CPU machines have a 90 MHz P54C, 32 MBytes of RAM and 258
MBytes of disk (we also have a dual 450 MHz Pentium II node and a few
200 MHz Pentium nodes).  There are two networks connecting the nodes of
the dual CPU cluster: the public and private (see
Figure~\ref{fig:cyclone_arch} for details). 

  The public network is a ``fat'' tree interconnect with two 8-port
Gigabit switches at the root and 12 12-port 100 Mbit switches at the
nodes, each connecting 12 PCs at the leaves.  The 100 Mbit switches are
connected to the Gigabit switches via fiber-optic cable; the PCs to the
100 Mbit switches via CAT 5 twisted-pair cable. 

  The private network contains 12 100 Mbit switches all connected to a
single 12-port 100 Mbit switch.  All connections in the private network
are CAT 5 twisted-pair cable.  The Sarnoff cluster can currently support
up to 192 PCs, with 100 Mbit node-to-node bandwidth. 

  The current cluster was designed and built in late 1997 for
approximately \$250K.  Today, the same amount of dollars would purchase
a cluster with double the compute power (dual 400 MHz Pentium II),
double the secondary memory (6 GB disks) and double the primary memory (128
MB RAM), and almost double the system bus (from 66 MHz to 100 MHz).

\section{The IDL Interpreted Language}

  There are many interpreted languages in wide use today.  Visual Basic
has O($10^6$) users worldwide while MATLAB, IDL, Mathematica and Maple
each have O($10^5$) users.  MP-NOW systems can take advantage of these
tools and their large user bases because they are built out of the same
components and use the same OSes interpreted languages are designed to
run on.  These languages provide many of the same features and mainly
differ in the applications of their primary users.  Visual Basic
predominates in the business arena, MATLAB is extremely common in the
signal processing community, IDL has been adopted for many satellite
image processing applications and Mathematica and Maple are used in
mathematical contexts.  Rather than go into the details of each of these
packages we have chosen to focus on IDL as a representative example.

  IDL (Interactive Data Language) from Research Systems, Inc. 
originally was focused on image processing, but now it includes
capabilities for building GUIs, doing 3D visualization and engaging in
fully object oriented programming.  Like most interpreted languages it
is very efficient at manipulating multi-dimensional arrays and provides
a rich notation for accessing and modifying these data structures.  For
example, the IDL vector statement: {\tt array[i:i+n] = cos(array[j:j+n])} is
fairly straightforward in its meaning: replace elements i through i+n
with the cosine of the values in elements j through j+n.  To perform a
similar function in a compiled language would require several lines of
code as well as the use of intermediary arrays which IDL creates and
uses implicitly.

  The price for this rich set of tools is that there are certain
specific types of calculations that do not run efficiently in
interpreted languages.  The implicit copying of arrays enables some of
the more powerful constructs of the these languages, but creating many
duplicates of large arrays can be prohibitive in very large memory
applications.  Interpreted languages are also inefficient at handling
large loop operations, which require the explicit generation and
evaluation of all the code in the loop.  In general, programmers are
encouraged to express loops in vector notation.

\begin{figure}[tbh]
\centerline{\psfig{figure=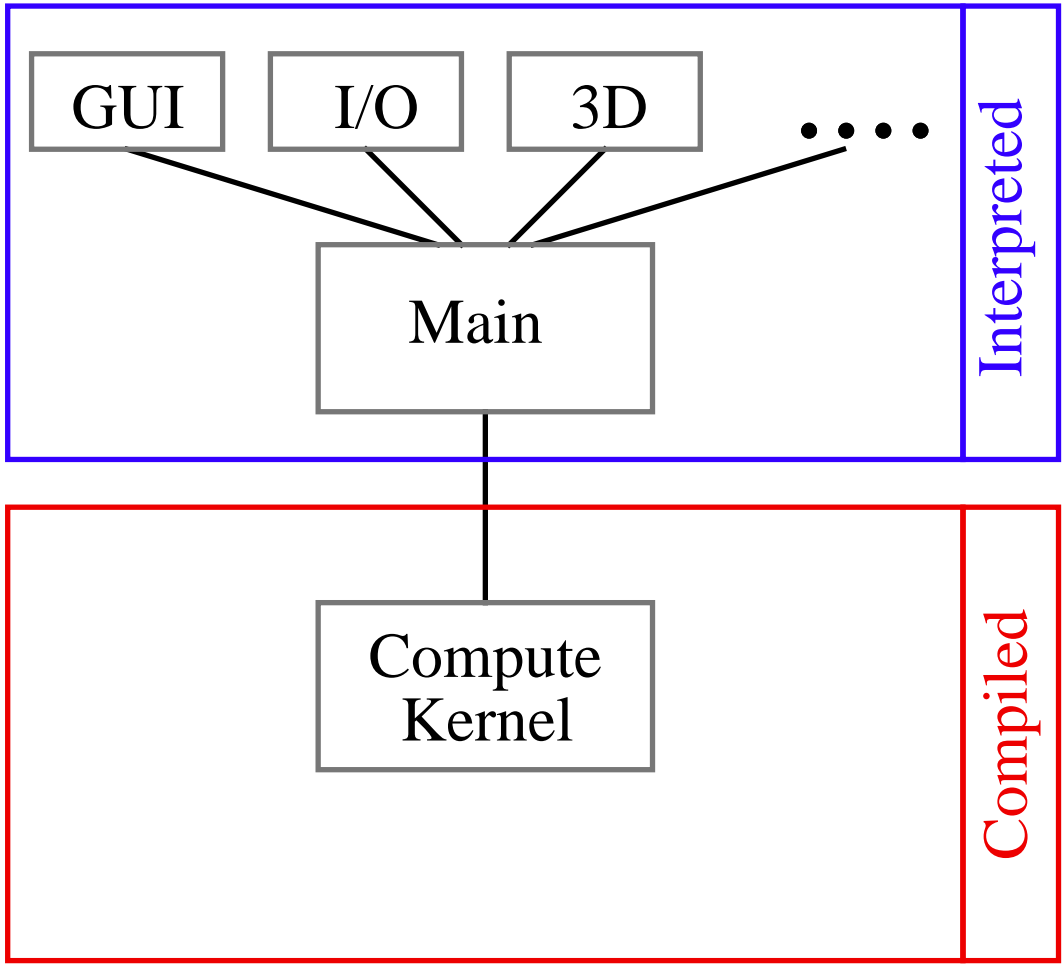}}
\caption{ {\bf Single CPU Application Architecture.}
Application architecture before implementation on an MP-NOW. 
GUI and other ``high level'' operations are written in the interpreted
layer, which calls the compute kernel written in a compiled language. 
}
\label{fig:one_cpu_app_arch}
\end{figure}

  There are applications where loops or very large arrays are unavoidable,
in which case most interpreted languages provide a simple interface for
linking in compiled subroutines (written in C, C++ or Fortran).  This
allows the programmer to take full advantage of an interpreted language
for developing the parts of an application that are generally most time
consuming in a compiled language (e.g., GUI, I/O, and graphics) while
still being able to use a compiled language where performance
constraints require it.  This simple model for enhancing the performance
of interpreted languages is widely used and is shown schematically in
Figure~\ref{fig:one_cpu_app_arch}. 

  In IDL and most interpreted languages the calling of externally
compiled routines requires three pieces: a wrapper function, the
external library function, and a special set of compilation
instructions.  To illustrate these three components consider a simple
program that adds two vectors together.  The first piece is the IDL
wrapper function which takes the IDL arrays as input and calls the
external library function.  The wrapper function shown below also
computes the number of elements and allocates the output array for
storing the result.  Calling an external function requires using the IDL
{\tt CALL\_EXTERNAL} function which takes as inputs the name of the
library, the specific library function, and the arguments to be passed. 

\begin{verbatim}
      FUNCTION add_arrays,a,b
        ;Get size of array.
        N = N_ELEMENTS(a)

        ;Make output array.
        c = a

        ;Call external function, pass data.
        flag = CALL_EXTERNAL("add_arrays.so","_add_arrays_",N,a,b,c)
      RETURN,c
      END
\end{verbatim}

  The second component is the compiled function itself.  Shown below
is the code necessary to add two arrays together and return
the result in a third array.  This is an ordinary C function except
that the pointers to the arguments are passed through the argv
array and need to be recast into their original types:

\begin{verbatim}
      float add_arrays(int argc,void *argv[])
      {
        /* Declare local names of inputs. */
        long  N;  float *a,*b,*c;  long i;

        /* Cast the pointers in argv to local names. */
        N = (long)(*(long *)argv[0]);
        a = (float *) argv[1];  b = (float *) argv[2];  c = (float *) argv[3];

        /* Add vectors. */
        for(i=0; i<N; i++) {
          c[i] = a[i] + b[i];
        }

        /* Return flag value. */
        return(1.0);
      }
\end{verbatim}

  The third and final component consists of the special compilation
instructions that enable the library function to be used by IDL.  This
is usually the one architecture/OS dependent aspect of calling external
functions.  As an example, the instructions for compiling the {\tt
add\_arrays} subroutine within the Linux environment are:

\begin{verbatim}
      cc add_arrays.c -fPIC -o add_arrays.o -c add_arrays.c
      ld -shared -o add_arrays.so add_arrays.o
\end{verbatim}

  Although the details of this process for other interpreted languages
differ, the overall approach is the same.  Once constructed, this method
provides a seamless interface between the interpreted and compiled
environments and each is completely independent of the other. 

  Our approach to using interpreted languages on a MP-NOW is to take
advantage of this ability of most interpreted languages to call
externally compiled subroutines.  We break our application into an
interpreted layer and a compiled layer.  In the interpreted layer all
the ``code intensive'' operations are performed (i.e., GUI, I/O, program
management) while in the compiled layer resides a computational kernel
which does the CPU intensive component.  Within the compiled layer we
also insert the tools necessary to manage multiple instances of the
computational kernel on a MP-NOW.  This approach to implementing
parallelism places specific constraints on the type of tool than can be
used.  Specifically, the parallel layer must be callable from a compiled
subroutine.  In addition, the parallel tool cannot require a large
framework which cannot be imposed onto an existing interpreted language
which is unalterable by the user.  The TNT library described in the
next sections satisfies both of these constraints.

\section{The TNT Library}
    
\begin{figure}[tbh]
\centerline{\psfig{figure=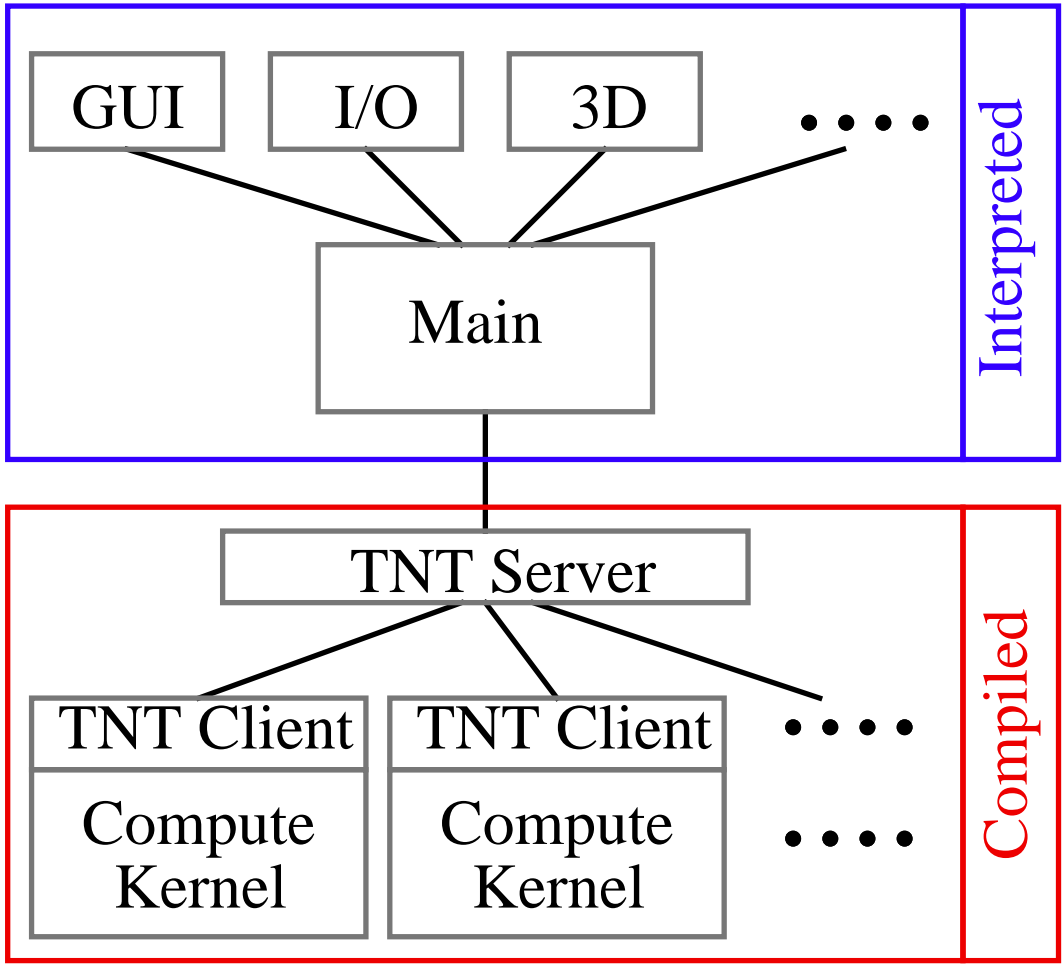}}
\caption{ {\bf MP-NOW Application Architecture.}
Application architecture after implementation on an MP-NOW. 
An additional ``TNT'' layer has been added to the compute kernel which
invokes and manages multiple copies of the compute kernel on a MP-NOW. 
}
\label{fig:mpnow_app_arch}
\end{figure}

  The Next generation Taskbag (TNT) is a client-server based
Applications Programming Interface (API) framework for distributing and
managing multiple tasks on a MP-NOW.  TNT is a C based library which can
be used in any compiled program.  As such, it is possible to insert the
appropriate TNT calls into the compiled layer called by an interpreted
language (see Figure~\ref{fig:mpnow_app_arch}). 

\begin{figure}[tbh]
\centerline{\psfig{figure=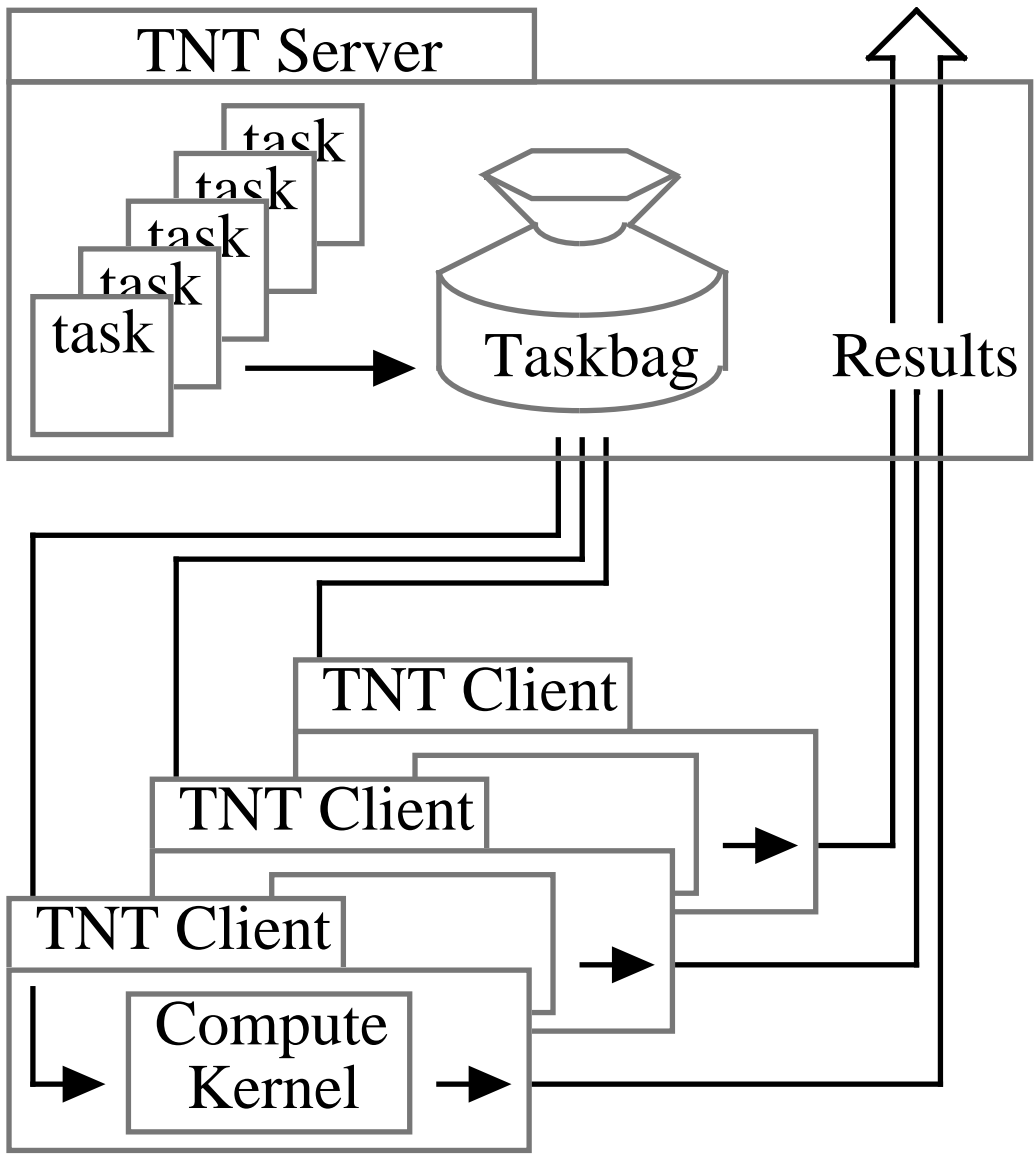}}
\caption{ {\bf TNT Application} A typical TNT application consists of a server
with many clients, communicating via TCP/IP.  The server: places tasks into
Taskbag; listens on a specific port for requests for tasks from clients;
dispatches tasks to requesting clients; accepts results from clients; monitors
status of clients and re-assigns tasks of dropped clients; when all tasks are
completed, returns results back to the main program.  The client(s) loop over
the Taskbag is until it is empty.  On each iteration a client will: send
requests for work to server on a specific port; read data sent by server over
network; call compute kernel with the data; send results of computation back
to server over network.}
\label{fig:tnt_app}
\end{figure}

\begin{figure}[tbh]
\centerline{\psfig{figure=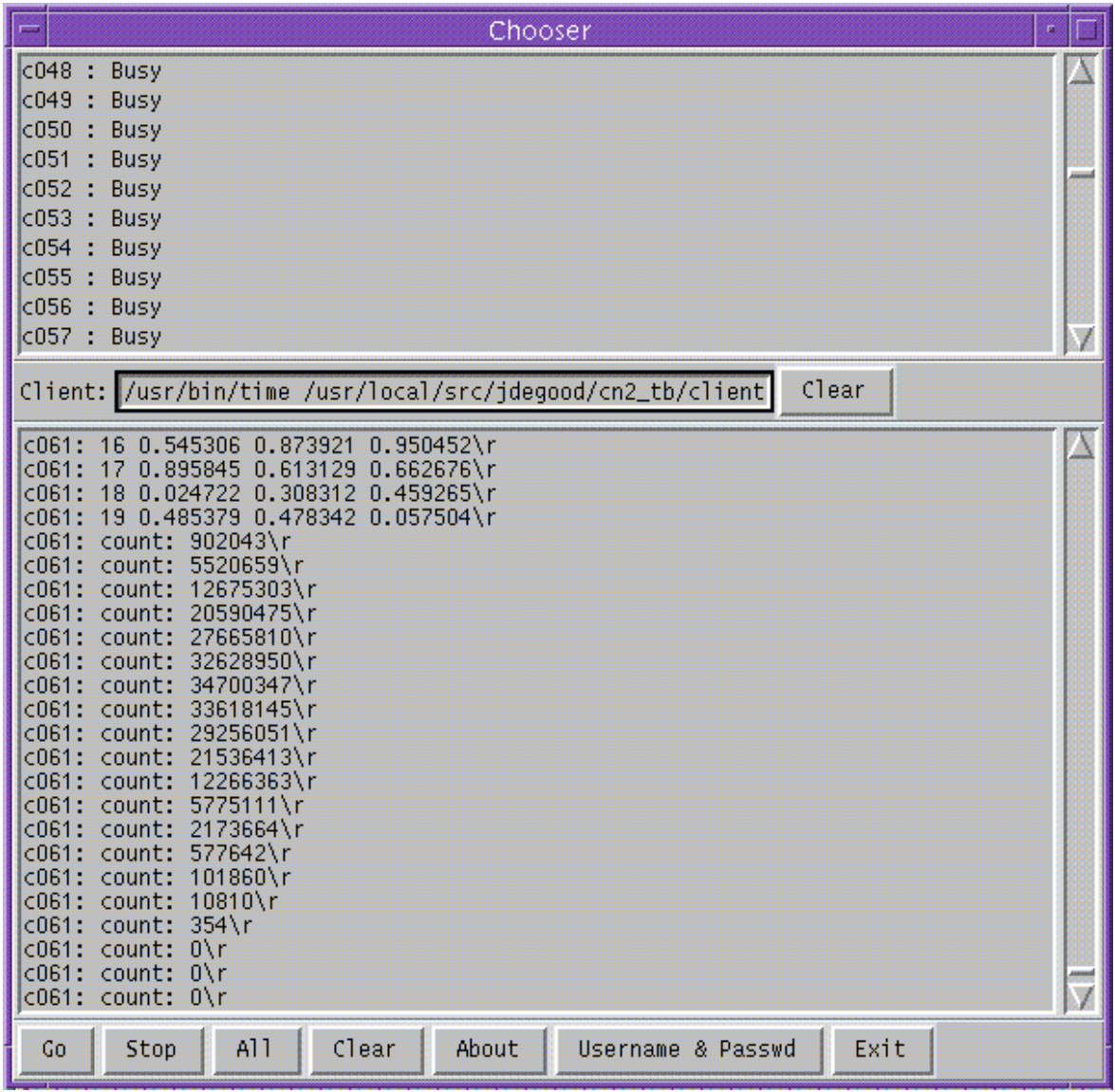,height=6.0in}}
\caption{ {\bf TNT CPU ``Chooser''.}  Interactive ``Chooser'' tool used by
applications programmer to select client nodes on a MP-NOW. The user chooses
what clients will execute the given command. In the case of the taskbag
application, the command would be an application-specific client executable. It
should be noted, however, that the client executable must reside locally on
each node or be NFS mounted on each node -- the chooser will not distribute the
executable.}
\label{fig:chooser}
\end{figure}

  The operation of a typical TNT application is shown in
Figure~\ref{fig:tnt_app}.  The server creates a ``Taskbag'' of work for
clients.  The clients are then executed remotely on a number of
processors.  The processors that the program is run on can be specified
either automatically or interactively via a ``Chooser'' (see
Figure~\ref{fig:chooser}).  The clients connect with the server and
request a task or taskbag (a group of tasks).  When they have completed
their tasks they return the results back to the server and ask for more
tasks. 

  The TNT library was developed on Linux (RedHat 5.0) and tested on
FreeBSD, NetBSD, and Solaris.  The entire library is written in C using
TCP sockets for server-client communication.  Server communicate with
the clients using ports, which allows simultaneous servers to be active
and listening to different port numbers. The library has many features,
including: multiple server support, interactive server, and inherent
load balancing.  These features are detailed below. 

  In TNT, a server can call client functions and a client can call
server functions.  This enables the creation of hierarchies of servers. 
For example, a ``root'' server can partition a large taskbag into many
sub-taskbags and distribute them to a collection of sub-servers.  These
sub-servers will then distribute tasks to the clients.  This allows for
a more efficient distribution of work across the cluster nodes. 

  Additionally, the server is interactive.  Current commands include:
{\it status}, {\it clients} and {\it quit}.  The {\it status} command
displays the clients connected to the server and what task each client
is currently working on.  The {\it clients} command displays all of the
client hostname and socket number pairs.  Finally, {\it quit} gracefully
terminates the server by allowing the clients to finish the current task
but not allowing additional tasks to be given out.  Once all clients
have returned the current task, the server closes.  In a future version
of TNT, we will allow a customizable server command set (implemented
with the Tcl library).  The API will have functions that allow the
programmer to specify additional commands (or modifications to intrinsic
commands). 

  Finally, TNT is inherently load balancing in the sense that when a
client finishes a task it requests additional work. If there are no
tasks remaining then the client exits and frees up the processor.  The
processors that run faster will pick up more work and slower processors
will pick up less work.  There is no static assignment of work to
processors.

\subsection{TNT API Overview}
  The available functions in the TNT library are grouped into two
categories: server and client.  Next we provide a brief overview of
available function classes for each group. 

\subsubsection{Server Functions}
\begin{description}
 \item [Network Initialization] Network initialization sets the server to allow
	client connections on a specified network port.

 \item [Registry] The server maintains a list of active clients.  This list is
	called the registry.  The server may add clients, remove clients and
	verify clients.  These functions are usually called in response to a
	client request for addition, removal or verification.

 \item [Taskbag Operations] The server can add and remove tasks from the
	taskbag. There are also functions for locating tasks in the taskbag.

\end{description}

\subsubsection{Client Functions}
\begin{description}
 \item [Server Connection] Given a hostname and port, the client calls a
 function to connect to the server.  This sets up a persistent socket
 connection with the given server.

 \item [Registration] After the client has established a connection with the
 server, it requests registration from the server.  The server will respond
 with success or failure. A failure indicates that the client is already
 registered or invalid parameters were provided.

 \item [Request Task (or Taskbag)] Once the client is registered with the
 server, it will typically enter a request task loop.  In this loop, the client
 will return the previous, completed task (null, if initial task) and request a
 new task.

 \item [Close Session] Once the server indicates that there are no more tasks,
 the client calls a function to close the current session.  This will cause the
 server to remove the client from the registry and close the socket connection.

\end{description}

\subsection{TNT Essentials}
The essentials of the TNT API can be summarized as follows:

\begin{itemize}
  \item[] \hrulefill
  \item[] \centerline{\large \bf Basic TNT API}
  \begin{itemize}
    \item[] -- TNT client/server templates contain calls to TNT library.
    \item[] -- Programmer replaces default functions in templates.
    \item[] -- Typical application requires programmer to write the following
               functions:
  \end{itemize}
  \item[] \centerline{\large \it Server}
  \begin{itemize}
    \item[] {\tt ApplicationServerInit()}
      passes data from the interpreted layer to the TNT server.
    \item[] {\tt FillTaskbag()}
      passes data to the clients by placing tasks in Taskbag.
    \item[] {\tt PrintTaskbagResults()}
       returns result of computation back to Main.
  \end{itemize}
\item[] \centerline{\large \it Client}
  \begin{itemize}
    \item[] {\tt ApplicationClientInit()}
      performs any client initialization.
    \item[] {\tt ProcessTask()}
      calls the unmodified computational kernel.
  \end{itemize}
\item[] \centerline{\large \it Server and Client}
\begin{itemize}
    \item[] {\tt InitCmdLineVars()}
      initializes any user-configurable application variables.
    \item[] {\tt ParseCmdLine()}
      parses command-line arguments and sets any user-configurable variables.
  \end{itemize}
  \item[] \hrulefill
\end{itemize}

The TNT API consists of template server and client programs each
containing several function calls, which are customized by the
programmer to a particular application.  In a typical application, three
server functions {\tt ApplicationServerInit()}, {\tt FillTaskbag()} and
{\tt PrintTaskbagResults()}; two client functions {\tt
ApplicationClientInit()}, {\tt ProcessTask()} and two command-line
argument functions {\tt InitCmdLineVars()}, {\tt ParseCmdLine()} are
modified.  {\tt ApplicationServerInit()} performs any server-specific
initialization, {\tt FillTaskbag()} passes data from the server to the
clients, {\tt PrintTaskbagResults()} sends results of computation back
to the main program, and {\tt ProcessTask()} calls the unmodified
computational kernel. 

\subsection{Overhead}

  One of the essential benefits of using the TNT library with a compiled
language is the very low overhead it introduces at the coding, system,
compute and communication levels.

  The primary coding overhead occurs in creating the interface between
the interpreted language and the computational kernel (see section 3). 
If a compiled computational kernel is required because the nature of the
calculation does not lend itself to an interpreted language, then this
overhead will have already been incurred.  However, if the computational
kernel is readily programmed in an interpreted language, then the
overhead of converting the kernel to a compiled language and interfacing
is required in order to subsequently take advantage of the TNT tools.
A secondary amount of coding overhead is necessary to insert the TNT
client/server framework.  This consists of creating the appropriate
wrapper functions as well as packing and unpacking the data into ``tasks''
by the server and client.

  The TNT library is built using TCP sockets and does not require any
other system resources (e.g. daemons, shell programs, etc...) other than
those necessary to launch the interpreted language and the clients.
After launch the computational kernel is executed via an
external library call, which induces no more overhead than that required
for other library calls.  The computational kernel starts the server,
and may start the clients.  The clients can be launched interactively or
directly by the application through remote shell commands.  In cases
where very rapid launching is required (e.g. applications with a
real-time constraint) higher performance multi-launch mechanisms can be
used \cite{VEX}.

  The computational overhead of this approach is limited to the
overhead of packing the data into tasks on the server and unpacking
the data on the client side.  Additional overhead might
be required to divide the work up into tasks (e.g., computing subsets
of indices) and for pooling the results returned by each client.

  As with any mechanism for distributing work across a MP-NOW,
communication overhead is unavoidable and managing this overhead is
often critical to achieving good performance.  In addition, the
client/server model is strictly limited to independent,
non-communicating tasks. The primary communication overhead comes from
the server transferring the data associated with a given task to the
requesting client and any results that are sent back by the client to
the server.  If the amount of data being sent is large compared to the
subsequent computations being performed then communication time can
easily dominate.

\section{Task Parallel Applications}

  Task parallelism is one of the most common types of parallelism that
exist in a wide array of science and engineering applications.
Many of these applications also fit the model whereby a small
amount of code accounts for a majority of the computations and can benefit
from an approach that allows much of the code to be written in an
interpreted language. The TNT based approach presented here is
generally applicable to any task parallel problem.  An analysis
of even a small fraction of these applications is beyond the scope
of this paper.  However, it is possible to examine several applications
in the fields of Astronomy and Astrophysics to determine
the usefulness of the TNT based approach.

  Astronomy and Astrophysics are among the largest consumers at national
computing centers \cite{NCSAdata}.  These applications can be
grouped into two general classes: astrophysical simulations and
astronomical data analyses.  Within these categories, there are two
primary ways parallel computing resources are exploited: parameter space
studies (running the same ``small'' program many times with different
inputs) and large data intensive calculations that require multiple
computers to complete.

  The parameter space studies map onto a task parallel approach
trivially. Examples of this type of calculation are: calibration of
supernova models over a range of compositions
\cite{NERSC,Perlmutter98} to determine the size of the Universe;
calibration of star models over a range of ages, compositions and
masses to determine the age of the Universe; 
comparison of stellar oscillation data against hypothetical planetary
arrangements for finding extra-solar planets.

  Data intensive calculations are much more challenging to map onto a
task parallel model.  However, these types of calculations do stress the
limits of task parallelism.  Examples of the kind data intensive
calculations that can be mapped onto  task parallelism are particle
simulations, correlation analysis and pattern detection.

  Frequently in astrophysics it is desirable to simulate physical
systems as the interaction of many particles.  For example, to simulate
the motion of stars moving about the galaxy it is necessary to compute the
force of every star (particle) on every other star. A similar situation
occurs when the behavior of gaseous objects is modeled as a collection
of particles.  In this case the behavior of an individual particle is a
function of its nearest neighbors.

  Astronomical databases often consist of the positions of a large number
of objects.  One of the most common analyses performed on these data
sets is computing the correlation function of one type of object (e.g.
red stars) with another (e.g. blue stars).  The essence of the
correlation function is computing the relative distances every object.

  Another type of analysis that is performed on astronomical datasets is
pattern recognition or cluster detection.  In this case the dataset is
convolved with a filter for the type desired object, such as a high
spatial concentration of stars.  The essence of this operation is
finding all the objects that are near another object and using this
subset of data to evaluate the filter at that point.

  The common feature of the above data intensive problems is that they
use the same computational kernel.  This kernel takes as its input a
list of positions and returns as its output the distance of each point
from every other point or a list of the nearest points. Exploring
the performance of this one kernel provides information on the utility
of the TNT based approach to all the above examples.

\section{Implementation}

  In the previous section several problems in Astronomy and Astrophysics
were described with a common computational kernel.  This kernel has been
implemented using the TNT based approach described above.  The calculation is
set-up using code written in IDL which calls a computational kernel
written in C.

 More specifically, the computational kernel takes as its arguments
a set of N vectors ($\xb_1$,...,$\xb_i$,...$\xb_N$) each with $D$
elements.  The individual elements of the vectors can be made up of
real, complex, integer, string or mixed type data.  Within the kernel
there is a function called distance($\xb_i$,$\xb_j$) which returns a
positive real value corresponding to the separation of the vectors
$\xb_i$ and $\xb_j$ in the $D$ dimensional space.  The goal is to
compute the distance between every pair of vectors and to return a list
of the $M$ nearest neighbors to each point.

  The algorithm used to solve this problem is a simple direct
calculation that involves performing $N^2$ distance calculations and N
sorts each requiring O($N$~log~$N$) operations.  This method is not the
most efficient in all circumstance, however it is the most general and
works for all values of D and all distance functions.  This test problem
is simple to parallelize by putting $N/$\#CPU vectors on each processor.
Furthermore, by adjusting the parameters $N$ and \#CPU this problem can
readily probe both the computation dominated and communication dominated
regimes.

\section{Results}

\begin{figure}[tbh]
\centerline{\psfig{figure=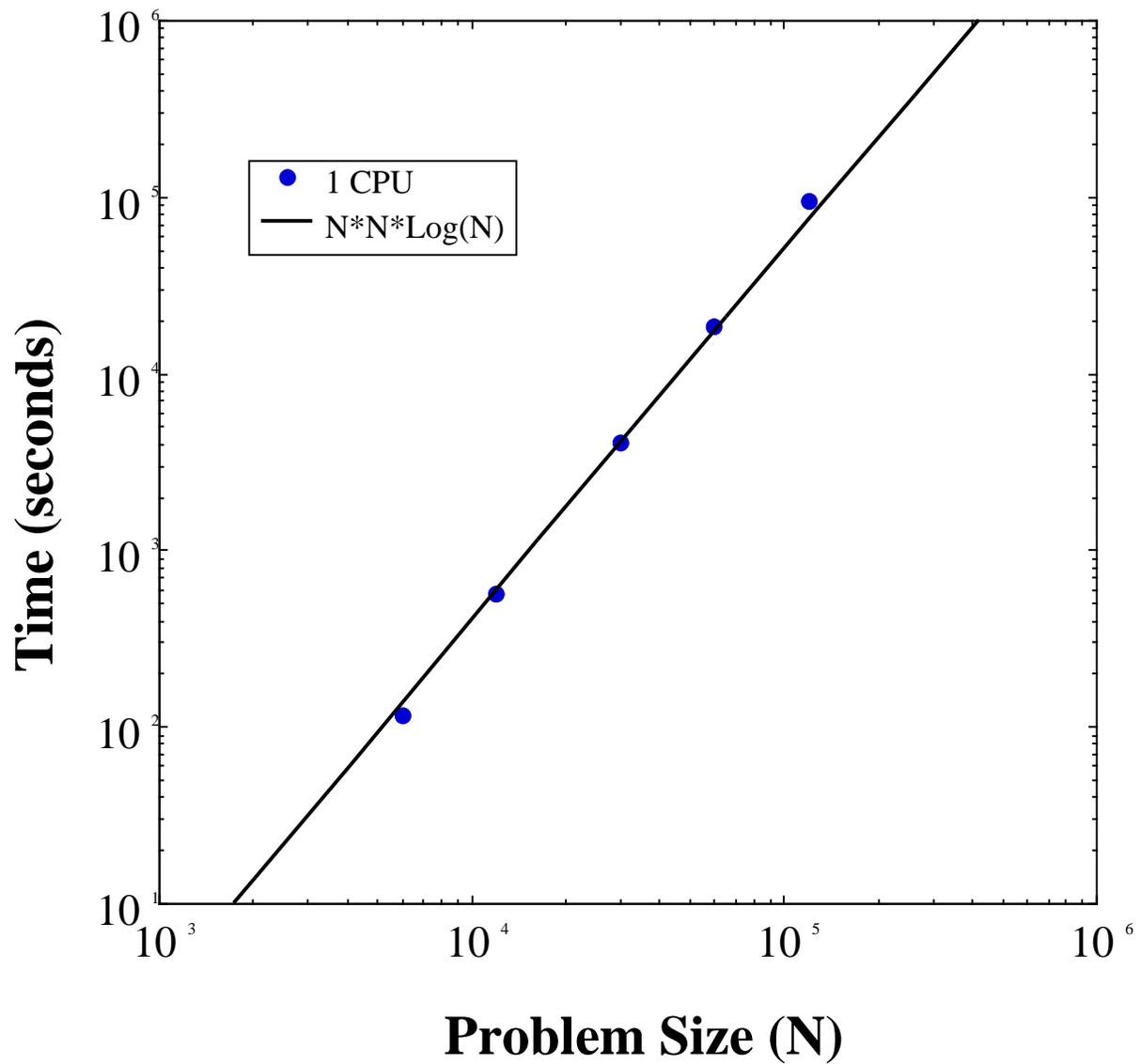,height=6.0in}}
\caption{ {\bf Problem Scaling.}
Single processor CPU times as a function of problem size for
the example pattern recognition problem.
}
\label{fig:one_cpu_times}
\end{figure}

  Our test problem has been implemented on the Sarnoff Cyclone.  The
execution times for various configurations of problem size and \#CPU are
shown in Table~\ref{tab:mpnow_M100_times}.  As a baseline, we look at
the single CPU behavior as a function of N. 
Figure~\ref{fig:one_cpu_times} shows that this problem scales in
the predicted $N^2$~log~$N$ fashion.  The same behavior is also exhibited
for larger numbers of CPUs.

\begin{figure}[tbh]
\centerline{\psfig{figure=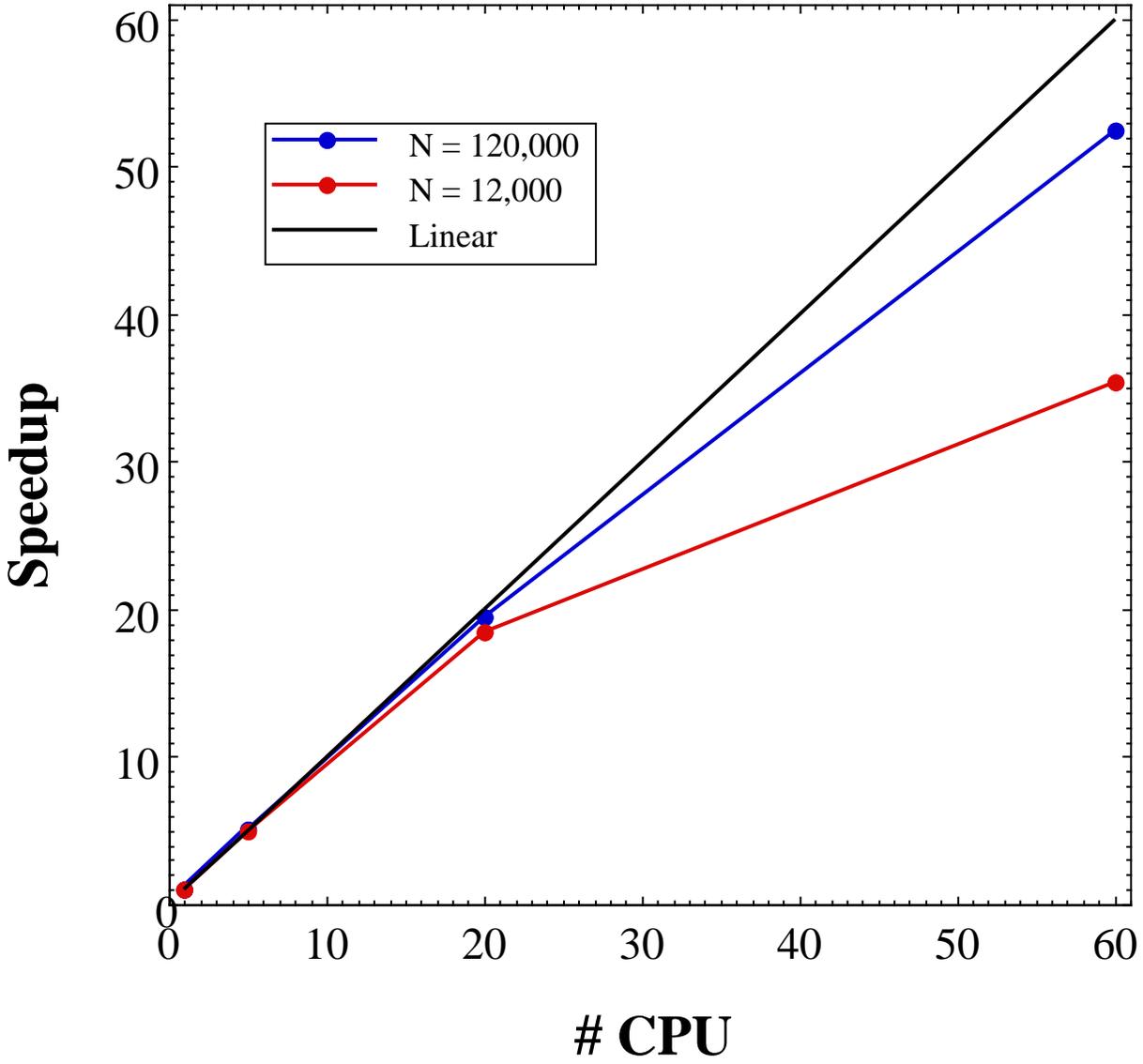,height=6.0in}}
\caption{ {\bf Parallel Performance.}
Speedup vs \#CPU for two different problem sizes.  The black line
indicates linear speedup.  The larger problem size shows nearly linear
speedup: a factor of 52 on 60 CPUs.  Task dispatch overhead is
evident on the smaller problem size. 
}
\label{fig:speedup}
\end{figure}

  The parallel performance results are shown in Figure~\ref{fig:speedup}
for two problem sizes using 1, 5, 20, and 60 CPUs.  Speedup of the
smaller problem is primarily constrained by the time required to
dispatch the parallel tasks, which becomes significant when the
computation time becomes on the order of a few seconds. The bigger
problem scaled much better, as the task dispatch overhead is much
smaller relative to the computation time.  The main limit to the speedup
of the bigger problem size are network communication between the client
CPUs and the TNT server.

\section{Conclusions and Future Work}

  With less than 3 weeks of total effort we were able to successfully
implement this paradigm on a readily obtainable MP-NOW system. 
Subsequent porting of similar applications should take only 1 or 2
days of effort.  The TNT API is simple and provides good performance,
resulting in speedups of greater than 50 on 60 CPUs.  Continued development
of the TNT library will broaden the types of problems that can be
addressed as well as increase the range of available OSes and
architectures.  As the Cyclone computer and TNT library are further
improved we expect to see the performance for this test problem
improve further. 

  TNT uses a client/server framework which naturally lends itself to
problems that exhibit a large amount of coarse grained parallelism.  In
addition, this model allows for dynamic load balancing as well as
rescheduling of non-completed tasks.  The client/server model is not
applicable to problems that require a large amount fine grained
parallelism or intertask communication.  For these types of
problems other tools have been developed for the Sarnoff Cyclone. 

  In the future we plan to extend the TNT library and to test it with
additional interpreted languages (e.g.  MATLAB).  We also plan to use
the current TNT/IDL implementation on several scientific data processing
applications, which will provide a further evaluation of the ease of use
and performance. 

\section*{Acknowledgments}
Jeremy Kepner would particular like to thank his Ph.D. advisor
Prof. David Spergel for supporting this work.

\clearpage
\newpage


\clearpage
\newpage


%

\begin{table}[tbh]
\begin{center}
\begin{tabular}{|l|ccccc|}
\hline
       & N = 6,000 & N = 12,000 & N = 30,000 & N = 60,000 & N = 120,000 \\
\hline
\#CPU = 1  & 116  & 535 & 4,080 & 18,600 & 94,000 \\
\#CPU = 5  & 24.5 & 107 & 826 & 3,990 & 18,600 \\
\#CPU = 20 & 7.8 & 29 & 213 & 1,010 & 4,820 \\
\#CPU = 60 &  5.5 & 15.1 & 78 & 356 & 1,790 \\
\hline
\end{tabular}
\end{center}
\caption{
MP-NOW execution times in seconds for various numbers of points (N)
and processors (\#CPU) and a constant number of neighbors (M = 100). 
}
\label{tab:mpnow_M100_times}
\end{table}


\clearpage
\newpage

\clearpage
\newpage

\section*{Biographies}
\begin{figure}[tbh]
{\bf Jeremy Kepner} received his B.A. in Astrophysics from Pomona
College (Claremont, CA).  He obtained his Ph.D. focused on Computational
Science from the Dept. of Astrophysics at Princeton University
in 1998, after which he joined MIT Lincoln Lab.  His research
has addressed the development of parallel algorithms and tools
and the application of massively parallel computing to a variety
of data intensive problems. 
E-mail:jvkepner@astro.princeton.edu or kepner@ll.mit.edu
\end{figure}

\begin{figure}[tbh]
{\bf Maya Gokhale} is a researcher in Space Data Systems at Los Alamos
National Laboratory. Previously she was head of the Advanced Computing 
and Networks Group at Sarnoff Corp.
Gokhale obtained her Ph.D. from the University of Pennsylvania in 1983.
Dr. Gokhale has spent the last ten years developing architectures
and tools for high performance computing systems. She was a principal member of
the Splash FPGA Computing team at the Supercomputing Research Center, where she
built CAD tools that were used to program Splash-1 and -2. 
Gokhale developed a
data parallel C compiler that targeted a novel Processor-in-Memory
array, Terasys (featured in an HPCC Blue Book) and its 
follow-on, the Cray3/SSS, as well as a cluster of workstations.
Gokhale leads several DARPA-sponsored projects in Adaptive Computing Systems.
Research thrusts include tools to map high level language onto configurable
hardware and to optimize the compilation of C to hardware circuits. 
Other on-going projects address distributing computation over clusters
of workstations.
Gokhale has numerous publications in
Adaptive and High Performance Computing. She is a reviewer for NSF and other 
government agencies and has served on the program committee and tutorials committee
of Supercomputing 'XY. 
E-mail: maya@lanl.gov
\end{figure}

\begin{figure}[tbh]
{\bf Ron Minnich} is on the research staff of the Advanced Computing Lab at Los
Alamos National Labs. His current research involves cluster computing,
high performance networking, adaptive network card architectures, and
operating systems support for distributed computing. Recent work includes
a process-private name space for Linux to support trusted, wide-area
distributed computing; the MINI ATM interface, which provides a Virtual
Interface Architecture model to programs; and the construction of the
Cyclone cluster at the Sarnoff Corporation. He received his Ph.D. at the
University of Pennsylvania in 1991.
E-mail: rminnich@acl.lanl.gov
\end{figure}

\begin{figure}[tbh]
{\bf Aaron Marks} received his B.S. and M.S. in Computer Science from East
Stroudsburg (East Stroudsburg, PA) and is currently working on his Ph.D. in
Computer and Information Science at the University of Pennsylvania
(Philadelphia, PA).  His research has focused on the design and construction of
MP-NOW systems and the design and implementation of distributed applications.
E-mail: amarks@sarnoff.com or ajmarks@seas.upenn.edu
\end{figure}

\begin{figure}[tbh]
{\bf John DeGood} received his B.S. in Chemistry from University of
Missouri-Rolla, and his M.S. in Computer and Information Sciences
from the University of Delaware.  His interests include real-time,
embedded, and system software.  Prior to joining Sarnoff
John developed analytical instrumentation at Hewlett-Packard.
E-mail: jdegood@sarnoff.com
\end{figure}

\end{document}